\documentclass[prl,aps,twocolumn,superscriptaddress,showpacs,epsfig]{revtex4}
\usepackage{graphicx}% Include figure files
\usepackage{dcolumn}% Align table columns on decimal point
\usepackage{bm}% bold math

\def\prn#1{{\left(#1\right)}}

\def\abrk#1{{\left\langle#1\right\rangle}}

\def\fig_width{3. in} % width of single column figure in PR

\def\prn#1{{\left(#1\right)}}
\def\brk#1{{\left[#1\right]}}

\def\abs#1{{\left|#1\right|}}

\begin{document}

\title{Measurement of the forbidden $\bm{6s^2 \, ^1S_0
\rightarrow 5d6s \; ^3 D_1}$ magnetic-dipole transition amplitude
in atomic ytterbium }
\author{J.~E.~Stalnaker}
\author{D.~Budker}

\affiliation{Department of Physics, University of California at
Berkeley, Berkeley, California 94720-7300}

\affiliation{Nuclear Science Division, Lawrence Berkeley National
Laboratory, Berkeley, California 94720}

\author{D.~P.~DeMille}

\affiliation{Department of Physics, Yale University, New Haven,
Connecticut 06520}

\author{S.~J.~Freedman}

\affiliation{Department of Physics, University of California at
Berkeley, Berkeley, California 94720-7300}

\affiliation{Nuclear Science Division, Lawrence Berkeley National
Laboratory, Berkeley, California 94720}

\author{V.~V.~Yashchuk}
\affiliation{Department of Physics, University of California at
Berkeley, Berkeley, California 94720-7300}
\date{\today}

\begin{abstract}
We report on a measurement of a highly forbidden magnetic-dipole
transition amplitude in ytterbium using the Stark-interference
technique.  This amplitude is important in interpreting a future
parity nonconservation experiment that exploits the same
transition.  We find $| \langle 5d6s \; ^3 D_1 \vert M1 \vert
6s^2 \, ^1S_0 \rangle | ~ = ~ 1.33(6)_{Stat}(20)_{\beta} \times
10^{-4} \mu_0$, where the larger uncertainty comes from the
previously measured vector transition polarizability $\beta$. The
$M1$ amplitude is small and should not limit the precision of the
parity nonconservation experiment.
\end{abstract}

\pacs{32.70.Cs,32.60.+i,32.80.Ys}

\maketitle

%%%%%%%%%%%%%%%%%%%%%%%   INTRODUCTION   %%%%%%%%%%%%%%%%%%%%%%%

The proposal to measure parity nonconservation (PNC) in the $6s^2
\, ^1S_0 \rightarrow 5d6s \, ^3D_1$ transition in atomic ytterbium
(Yb) \cite{demille95} has prompted both theoretical
\cite{porsev95,das97} and experimental \cite{bowers96,bowers99}
studies.  The magnetic-dipole ($M1$) amplitude for this transition
is a key quantity for evaluating the feasibility of a PNC-Stark
interference experiment as proposed in \cite{demille95}.  A
nonzero $M1$ amplitude coupled with imperfections in the
apparatus can lead to systematic uncertainties in a PNC
experiment. Here we present the first experimental determination
of the magnetic-dipole amplitude for the $^1S_0 \rightarrow
^3D_1$ transition. Our method is based on the technique of Stark
interference \cite{bouchiat74,chu77,gilbert84}.

\begin{figure}
\includegraphics[width=\fig_width]{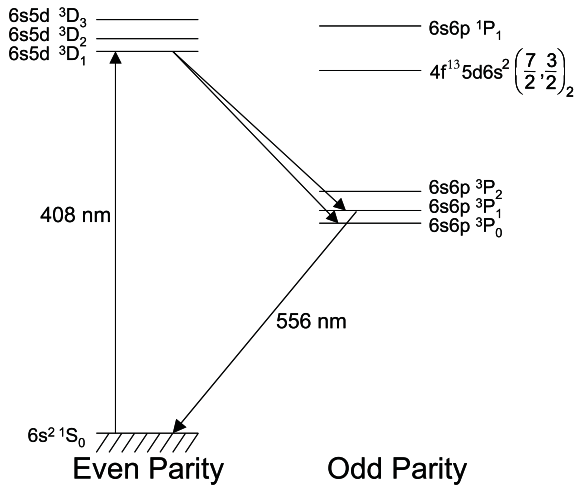}
\caption{Low-lying energy levels of Yb.} \label{energy}
\end{figure}

In the absence of external fields, the $^1S_0 \rightarrow ^3D_1$
transition (Fig. \ref{energy}) is highly suppressed.  An
electric-dipole transition amplitude $\brk{A\prn{E1}}$ is
forbidden except for parity mixing effects. A magnetic-dipole
transition amplitude is also forbidden because of the s-d nature
of the transition. Consequently, a nonzero transition amplitude
exists only as a result of configuration mixing and spin-orbit
interaction in both the upper and lower states \cite{demille95}.
There have been no detailed calculations of this amplitude.
Reference \cite{demille95} gives a rough estimate of $\vert
A\prn{M1} \vert \lesssim 10^{-4} \mu_0$, where $\mu_0$ is the
Bohr magneton.

%%%%%%%%%%%%%%%%%% STARK-INTERFERENCE TECHNIQUE %%%%%%%%%%%%%%

In the presence of a static external electric field, ${\mathbf
E}$, there is a parity-conserving mixing between the even parity
$^3D_1$ state and the odd parity states. For a \mbox{$J=0
\rightarrow J=1$} transition, this mixing leads to a Stark-induced
electric-dipole transition amplitude given by \cite{bouchiat74}
\begin{equation}
A\prn{E1_{St}} = i \, \beta \left( {\mathbf E} \times
\bm{\hat\varepsilon} \right) _{q}, \label{StarkAmp}
\end{equation}
where $\bm{\hat\varepsilon}$ is the direction of the polarization
of the laser light, $({\mathbf E} \times
\bm{\hat\varepsilon})_{q}$ is the $q$ component of the vector in
the spherical basis, and the vector transition polarizability
$\beta$ is a real parameter.  The magnitude of $\beta$ was
measured \cite{bowers99}:
\begin{equation}
|\beta| = 2.18(33) \times 10^{-8} {\rm ea_{0}/(V/cm)}.
\end{equation}

\begin{figure}
\includegraphics[width=\fig_width]{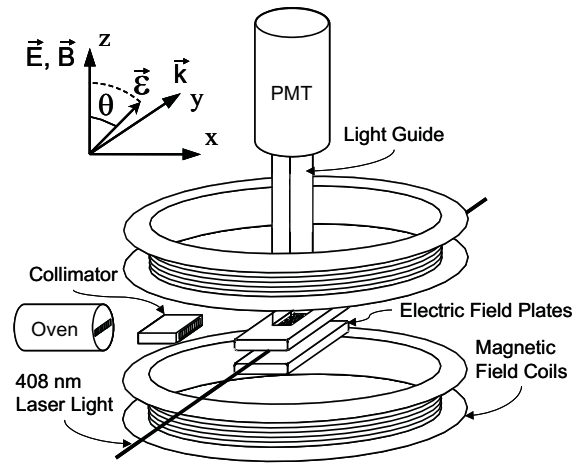}
\caption{Schematic of apparatus.} \label{expapp}
\end{figure}

In an electric field, the transition amplitude is the sum of the
Stark-induced $E1$ amplitude and the forbidden $M1$ amplitude.
The corresponding transition rate is
\begin{eqnarray}
W & \propto & \abs{A\prn{E1_{St}} + A\prn{M1}}^2 \\
  & \approx & \abs{A\prn{E1_{St}}}^2 + 2 {\rm Re}\brk{A\prn{E1_{St}}
  {A\prn{M1}^{\ast}}},\nonumber
\end{eqnarray}
where we neglect the contribution from $\abs{A\prn{M1}}^2$ since
$|A\prn{M1}| \ll |A\prn{E1_{St}}|$ with the electric fields and
polarization angles used here. The Stark-induced amplitude is
proportional to ${\bf E}$.  Thus, reversing ${\bf E}$ changes the
total transition rate, allowing the interference term to be
isolated from the larger terms. The $M1$ amplitude is given by
\begin{equation}
A(M1)= \langle ^3 D_1, \; M_J  \vert M1 \vert ^1S_0 \rangle ({\bf
\hat k } \times {\bf \hat\varepsilon})_{M_J},
\end{equation}
where ${\bf \hat k}$ is the direction of propagation of the
excitation light.  Equation \ref{StarkAmp} implies that only the
$M_J=\pm1$ components of the upper state are excited by
$A(E1_{St})$, where the axis of quantization is chosen along
${\bf E}$. With ${\bf k}$ perpendicular to ${\bf E}$, the sign of
the interference term is opposite for the transitions to the
$M_J=\pm1$ components, as can be verified by a simple
calculation. Thus, in order to observe the effect of the Stark-M1
interference, we apply a magnetic field, ${\bf B}$, allowing us
to resolve different magnetic sublevels. For ${\bf B}$ parallel
to ${\bf E}$ (Fig. \ref{expapp}) the interference term in the
transition probability is proportional to the rotational
invariant $( {\bf E} \times \bm{\hat\varepsilon} ) \times (
\bm{\hat k} \times \bm{\hat\varepsilon} ) \cdot \bm{\hat B}$.

Comparison of the difference in the transition rate between
opposite electric field states to the sum is a measure of the
fractional asymmetry $a$, defined as
\begin{eqnarray}
a & \equiv & {W(E_+)-W(E_-) \over W(E_+)+W(E_-)} \nonumber
\\  & = & {2 \langle ^3
D_1, \; M_J \vert M1 \vert ^1S_0 \rangle \over \beta E} {{\rm
cos}(\theta) \over {\rm sin}(\theta) } M_J,\label{asymEqn}
\end{eqnarray}
where $\theta$ is the angle between the dc electric field and the
polarization of the excitation light (Fig. \ref{expapp}). The
asymmetry increases with decreasing $\theta$ while the dominant
signal decreases as ${\rm sin}^2(\theta)$.  Most of the data was
taken at $\theta=\pm45^\circ$, where the interference term is
maximal.

%%%%%%%%%%%%%%%%   EXPERIMENTAL APPARATUS %%%%%%%%%%%%%%%%%%%%

Much of the apparatus used in this experiment had been used for
the measurement of the Stark-induced transition amplitude and is
described in detail in Refs. \cite{bowers99,bowers98}. A stainless
steel oven with a multi-channel nozzle created an effusive beam
of Yb atoms inside of a vacuum chamber with a residual pressure
of $\approx 5 \times 10^{-6}~{\rm Torr}$.  The oven nozzle
collimation resulted in a Doppler width for the $408$-${\rm nm}$
transition of $\approx~150~{\rm MHz}$. The oven was heated with
tantalum wire heaters operating at $\approx 500~^{\circ}{\rm C}$
in the rear with the front $\approx 100~^{\circ}{\rm C}$ hotter
to avoid clogging. Ytterbium has seven stable isotopes with both
zero and nonzero nuclear spin ($^{168}{\rm Yb}$, $^{170}{\rm
Yb}$, $^{172}{\rm Yb}$, $^{174}{\rm Yb}$, $^{176}{\rm Yb}$, $I=
0$; $ ^{171}{\rm Yb}$, $I = 1/2$; and $^{173}{\rm Yb}$, $I =
5/2$). To avoid significant overlap of the optical spectra of the
zero-nuclear-spin isotopes and the hyperfine components of the
nonzero-nuclear-spin isotopes, an external vane collimator was
installed; reducing the Doppler width to $\approx 15 ~ {\rm
MHz}$. The vane collimator was made by layering $0.076~{\rm mm}$
thick sheets of stainless steel foil between $0.94~{\rm mm}$
thick stainless steel spacers.  The length of the collimator was
$5.1~{\rm cm}$, providing a collimation angle of $\approx 1
^\circ $. The width of the collimator was $3.8~{\rm cm}$. The
collimator was heated using tantalum wire heaters to $\approx
350~^\circ {\rm C}$ to prevent clogging.  The collimator was
mounted on a movable platform, allowing precise alignment of the
angle of the collimator relative to the atomic beam during the
experiment. We estimate an atomic density of $\approx 2
~\times~10^9 ~{\rm cm}^{-3}$ in the interaction region.

Approximately $80 ~ {\rm mW}$ of laser light at $408 ~ {\rm nm}$
excited ytterbium atoms to the $^3D_1$ state in the geometry
shown in Fig. \ref{expapp}. The $408$-${\rm nm}$ light was
produced by frequency doubling of $\approx 1.2~{\rm W}$ of
$816$-${\rm nm}$ light from a titanium-sapphire laser (Coherent
899-21) pumped with $\approx 12 \; {\rm W}$ from a multi-line
argon-ion laser (Sectra Physics 2080). A commercial bow-tie
resonator with a Lithium-Triborate crystal (Laser Analytical
Systems Wavetrain cw) provided frequency doubling.

\begin{figure}
\includegraphics[width=\fig_width]{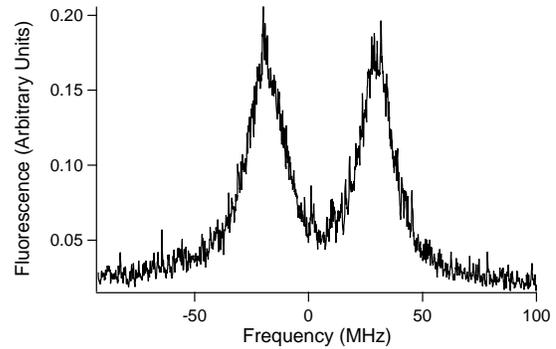}
\caption{Typical power-normalized, single scan over the $M_J~=~
\pm ~1$ components of the $6s^2 ~ ^1S_0 ~ \rightarrow 5d6s ~
^3D_1$ transition of $^{174}{\rm Yb}$ with $|{\bf E}|= 15~{\rm
kV/cm}$, $\theta=45 ^{\circ}$, $|{\bf B}|=36~ {\rm G}$.}
\label{scan}
\end{figure}

A Burle 8850 photomultiplier tube (PMT) monitored the
fluorescence in the $6s6p \; ^3 P_1 \rightarrow 6s^2 \; ^1S_0$
decay channel at $556 ~ {\rm nm}$ (Fig. \ref{energy}). The top
electrode had an array of $198$ $0.12~{\rm cm}$ diameter holes,
allowing the fluorescence to be collected by a Lucite light guide
and conducted to the PMT.  The presence of the holes in the top
field plate reduced the electric field between the plates. This
effect was calculated to be less than $1\%$ using a random walk
solution to Laplace's equation \cite{buslenko66}.  An
interference filter with transmission centered at $560~{\rm nm}$
with a $10~{\rm nm}$ full width at half maximum was placed in
front of the photomultiplier tube in order to limit detection of
scattered light at $408~{\rm nm}$. Approximately $\approx 0.05\%$
of the atoms undergoing a transition were detected, resulting in
typical photocathode currents of $10^4 ~ {\rm e/s}$ on the peak
of the resonance for $|{\bf E}|=15~{\rm kV/cm}$.  The
off-resonance background was consistent with the expected PMT
dark current and residual scattered light; contributing a noise
$\approx 3$ times less than the signal shot noise for $|{\bf
E}|=15~{\rm kV/cm}$. The laser frequency was scanned $\approx 200
~ {\rm MHz}$ with both increasing and decreasing frequency over
the transition and the fluorescence spectrum was recorded with a
digital oscilloscope.  The scan time each way was typically $1 ~
{\rm s}$. A typical single scan is shown in Fig. \ref{scan}.

After the laser was scanned the polarity of the electric field
was either switched or left unchanged in accordance to the
following pattern: $\; \left( + ~ - ~ - ~ + ~ - ~ + ~ + ~ -
\right)$.  This pattern was chosen to limit systematic effects
associated with drifts in the laser frequency and atomic beam
intensity. A bipolar power supply (Spellman CZE1000R), modified
so that the polarity was computer controlled, produced the high
voltage used in the experiment. The polarity of the top electrode
was reversed while the bottom electrode remained grounded.  A
resistor divider monitored electrode voltage. The magnitude of
this voltage changed by $< 0.1\%$ with the change in polarity. A
delay of $\approx 2~{\rm s}$ after each switch allowed the
electric field plates to fully charge before the next scan. The
stainless steel electric field plates were separated $1.016~{\rm
cm}$ by Delrin spacers.  The typical value of the electric field
was $\approx 12~{\rm kV/cm}$. After each sequence of E-field
switches, the polarity of the magnetic field was switched
according to the $\; \left( + \; - \; - \; + \right)$ pattern.
The Earth's magnetic field was reduced to $\lesssim 50 ~{\rm mG}$
with external coils.  A pair of in-vacuum coils in a near
Helmholtz configuration provided the uniform magnetic field
needed for the experiment. A typical magnetic field was $36~{\rm
G}$. A run consisted of $5$-$10$ sets of $32$ forward and
backward laser scans ($8$ E-field switches $\times$ $4$ B-field
switches per set) with fixed values of $\theta$, $|E|$, and
$|B|$. Periodically $\approx 50$ laser scans were taken with zero
magnetic field in order to monitor for changes in the lineshape
of the transition due to temperature fluctuations of the oven and
collimator.

A temperature stabilized, hermetically sealed Fabry-Perot cavity
with a free spectral range of $150 ~ {\rm MHz}$ and finesse of $
\approx 15$ was used to monitor a portion of the $816$-${\rm nm}$
light. A photodiode monitored the $408$-${\rm nm}$ laser power in
order to normalize the signal for power fluctuations. The
transmission through the Fabry-Perot and the $408$-${\rm nm}$
laser power were recorded concurrently with the fluorescence
signal.

%%%%%%%%%%%%%%%  ANALYSIS %%%%%%%%%%%%%%%%%%%%%

The Fabry-Perot transmission peaks were used to line up the
fluorescence spectra of two consecutive laser scans in order to
eliminate frequency drift between scans.  Two scans at opposite
electric fields and the same magnetic field were combined.  The
sum of the two fluorescence spectra was fit to the function
\begin{eqnarray}
W(E_+)+ W(E_-) & = & \nonumber \\ \zeta ( f \left( \nu - c_1
\right) & + & f \left( \nu - c_2 \right) ) + m \; \nu + b,
\end{eqnarray}
where $\nu$ is the frequency of the laser, $c_{1 \left( 2
\right)}$ is the center position of the first (second) peak,
$\zeta$ is the amplitude of the peaks, $m$ and $b$ account for any
linear background coming from scattered light.  The function $f$
was numerically determined from the spectra taken at zero magnetic
field. Because the sign of the interference term is opposite for
the different magnetic sublevels (see Eq. \ref{asymEqn}), the
spectral dependence of the asymmetry is given by the difference
between the $M_J= \pm 1$ peaks multiplied by an asymmetry
coefficient. The difference between the two spectra was therefore
fit to a function whose line shape was constrained by the fit
parameters from the sum fit:
\begin{eqnarray}
W(E_+)-W(E_-) & = & \nonumber \\ a \; \zeta (  \; f \left( \nu -
c_1 \right) & - & f \left( \nu - c_2 \right) ) + C,
\end{eqnarray}
where $a$ is the asymmetry coefficient given in Eq.
\ref{asymEqn}, and $C$ accounts, at lowest order, for any
possible background that may be present in the difference due to a
constant offset in the electric field which does not change sign
with the electric field switch.  Higher order contributions to
the line shape from a constant offset in the electric field were
analyzed and found to be insignificant in the determination of
$a$. Changing the polarity of the magnetic field reverses the
sign of $a$ since the resonance frequencies of the magnetic
sublevels switch (Eq. \ref{asymEqn}). Note that the asymmetry
coefficient does not depend on the value of $|{\bf B}|$.

%%%%%%%%%%%%%%%%%%%%  RESULTS %%%%%%%%%%%%%%%%%%%%%%%

The measurements were performed on isotope $^{174}{\rm Yb}$ which
has a large relative abundance and is spectrally well isolated.
The $M1$ transition amplitude was measured in a variety of
different field values and configurations.  The variation of $E$
was from $5 \; {\rm kV/cm}$ to $20 \; {\rm kV/cm}$, $\theta$ from
$-70^{\circ}$ to $70^{\circ}$, and $B$ from $12 \; {\rm G}$ to $84
\; {\rm G}$. In addition, data was taken without the external
collimator in order to determine possible systematic effects
associated with the line-shape modeling or frequency noise.  For
this data the overlap between the $^{174}{\rm Yb}$ and
$^{173}{\rm Yb} \left( F={5 \over 2} \rightarrow F'={5 \over 2}
\right)$ lines was significant and the analysis modified to
include effects of this transition.

The effects of misalignments of the fields and imperfect
reversals were analyzed analytically and numerically using density
matrix formalism. These calculations indicate that systematic
effects are significantly smaller than the statistical
uncertainty.  Possible systematic effects are also severely
constrained by confirming the characteristics of the asymmetry.
The method of analysis described above is sensitive to
asymmetries which reverse sign with ${\bf E}$ and are of opposite
sign for the two magnetic sublevels. Equation \ref{asymEqn}
implies that the sign of the asymmetry should also reverse with
${\bf B}$ and $\theta$. Asymmetries which did not reverse sign
with either ${\bf B}$ or $\theta$ were consistent with zero. In
addition, the dependence of the magnitude of the asymmetry on the
magnitude of ${\bf E}$ and $\theta$ was also verified.

\begin{figure}
\includegraphics[width=\fig_width]{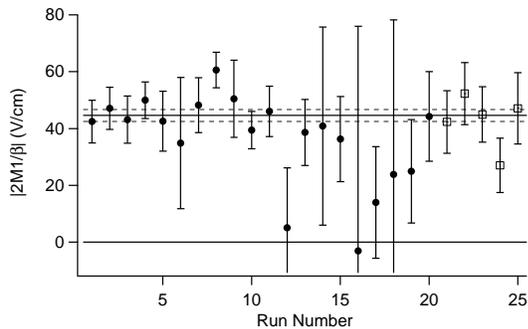} \caption{Experimental values of
$\abs{{2M1 \over \beta}}$.  The variation of $E$ was from $5 \;
{\rm kV/cm}$ to $20 \; {\rm kV/cm}$ and $B$ from $12 \; {\rm G}$
to $84 \; {\rm G}$. The circles (squares) represent data taken
with (without) the external collimator.  The solid line is the
mean and the dashed lines are the statistical error on the mean.
} \label{data}
\end{figure}

The final value of the $M1$ amplitude is based on data taken on
two different days. The data is shown in Fig. \ref{data}. The
statistical error for each point was estimated from the spread of
values obtained for each complete sequence of electric and
magnetic field switches within a given configuration.  These
errors are consistent with the expected limit due to shot noise.
Because the errors are estimated from the spread of $5$-$10$ sets,
there is some statistical variation in the size of the error
assigned to each run.  There is additional variation of the errors
due to differences in sensitivity for different polarization
angles (see Eq. \ref{asymEqn}) and differences in the amount of
data taken in a given configuration.  The final result is
\begin{equation}
{ 2 \langle ^3 D_1, \; M_J=\pm1  \vert M1 \vert ^1S_0 \rangle
\over \beta }  = - 44.6(21)_{Stat} \; {\rm V/cm}.
\end{equation}
This corresponds to an $M1$ amplitude of
\begin{equation}
| \abrk{^3D_1, M_J=\pm 1 | M1 | ^1S_0} | =
1.33(6)_{Stat}(20)_{\beta} \times 10^{-4} ~ \mu_0,
\end{equation}
where the second error represents the uncertainty in the
determination of $\beta$.

\begin{figure}
\includegraphics[width=\fig_width]{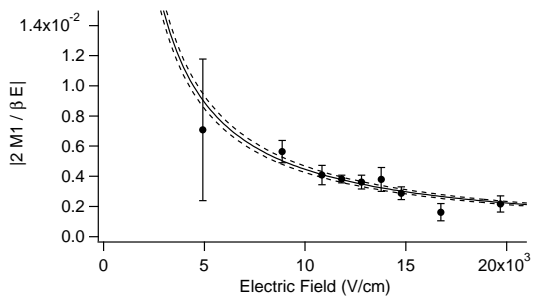} \caption{Measured values of
$\abs{{2M1 \over \beta E}}$ as a function of $|{\bf E}|$.  The
solid line is the expected dependence from the overall mean and
the dashed lines correspond to the errors on the mean.}
\label{eFieldPic}
\end{figure}

The measured value of $A(M1)$ agrees with the estimate in Ref.
\cite{demille95}.  This value is $\approx 3$ times larger than
the corresponding amplitude in the cesium (Cs) transition where
PNC is studied \cite{hoffnagle81,gilbert84,wood97,guena98}.
However, the expected large enhancement of the PNC amplitude in
Yb ($\approx 100$ times larger than in Cs \cite{demille95}) makes
the relative size of $A(M1)$ to the PNC amplitude smaller than it
is in Cs. An additional suppression of spurious interference
between the $M1$ amplitude and the Stark-induced amplitude is
possible by using the geometry for the PNC experiment employed in
Ref. \cite{drell85}. The reported measurement is for the isotopes
with zero nuclear spin. The isotopes with nonzero nuclear spin
($^{171}{\rm Yb}$, $I=1/2$ and $^{173}{\rm Yb}$, $I=5/2$) have an
additional contribution to the $M1$ amplitude and a small $E2$
amplitude due to hyperfine mixing effects. However, these
contributions are estimated to have values $\lesssim 10^{-5}
\mu_0$ \cite{demille95}, and should only lead to small
modifications to the present result. These effects will be
investigated in future work.  The size of the $M1$ amplitude
should not limit the precision of a Yb PNC measurement which is
in progress in our laboratory.

The authors thank M. Zolotorev and P. A. Vetter for many useful
discussions throughout this work and A. Vaynberg for help in
constructing the apparatus.  E. D. Commins and C. J. Bowers made
important contributions to early stages of the work. This work
was supported by the NSF, grant $PHY-9877046$.

%%%%%%%%%%%%%%%%%%%%%%%%%%%%%%%%%%%%%%%%%%%%%%%%%%%%%%%%%%%%%%%%%%%
%  References
%%%%%%%%%%%%%%%%%%%%%%%%%%%%%%%%%%%%%%%%%%%%%%%%%%%%%%%%%%%%%%%%%%%

%%%%%%%%%%%%%%%%%%%%%%%%%%%%%%%%%%%%%%%%%%%%%%%%%%%%%%%%%%%%%%%%%%%
%  Figures
%%%%%%%%%%%%%%%%%%%%%%%%%%%%%%%%%%%%%%%%%%%%%%%%%%%%%%%%%%%%%%%%%%%

\end{document}